# Feedback from nature: an optimal distributed algorithm for maximal independent set selection

Alex Scott
Mathematical Institute
University of Oxford, UK
scott@maths.ox.ac.uk

Peter Jeavons
Dept. of Computer Science
University of Oxford, UK
peter.jeavons@cs.ox.ac.uk

Lei Xu
Dept. of Computer Science
University of Oxford, UK
lei.xu@cs.ox.ac.uk

## ABSTRACT
Maximal Independent Set selection is a fundamental problem in distributed computing. A novel probabilistic algorithm for this problem has recently been proposed by Afek et al, inspired by the study of the way that developing cells in the fly become specialised. The algorithm they propose is simple and robust, but not as efficient as previous approaches: the expected time complexity is $O(\log^2 n)$. Here we first show that the approach of Afek et al cannot achieve better efficiency than this across all networks, no matter how the probability values are chosen. However, we then propose a new algorithm that incorporates another important feature of the biological system: adapting the probabilities used at each node based on local feedback from neighbouring nodes. Our new algorithm retains all the advantages of simplicity and robustness, but also achieves the optimal efficiency of $O(\log n)$ expected time.

## 1. INTRODUCTION
One of the most fundamental problems in distributed computing is to choose a set of local leaders in a network of connected processors so that every processor is either a leader or connected to a leader, and no two leaders are connected to each other. This problem is known as the maximal independent set (MIS) selection problem and has been extensively studied [16, 15, 3, 13, 14, 19, 18].

Different maximal independent sets for the same network can vary greatly in size. In contrast to the MIS selection problem, the related problem of finding a *maximum size* independent set (MaxIS) is notoriously hard. It is equivalent to finding a maximum clique in the complementary network, and is therefore **NP**-hard [12] (and hard to approximate [10]). However, computing an arbitrary MIS (not necessarily of the maximum possible size) using a centralised sequential algorithm is trivial: simply scan the nodes in arbitrary order. If a node $u$ does not violate independence, add $u$ to the MIS. If $u$ violates independence, discard it. Hence the real challenge is to compute such an MIS efficiently in a distributed way with no centralised control. Here we present a new distributed approach to this problem which achieves optimal efficiency.

Afek *et al.* have recently pointed out the similarity between the MIS selection problem and neural precursor selection during the development of the nervous system of the fruit fly *Drosophila* [2]. During development, certain cells in the pre-neural clusters of the fly specialise to become sensory organ precursor (SOP) cells, which later develop into cells attached to small bristles (microchaetes) on the fly that are used to sense the environment. During the first stage of this developmental process each cell either becomes an SOP or a neighbour of an SOP, and no two SOPs are neighbours. These observed conditions are identical to the formal requirements in the maximal independent set selection problem (see Figure 1).

Afek *et al.* also pointed out that the method used by the fly to select the SOPs appears to be rather different from standard known algorithms for choosing an MIS [16, 3, 18]. These algorithms rely on arithmetic calculations and precise numerical comparisons, and generally require explicit information about the number of active neighbours that each node in the network currently has. They also generally rely on exchanging complex messages representing precise numerical information. By contrast, the cells of the fly appear to solve the problem without clear central control using only simple local interactions between certain membrane-bound proteins, notably the proteins Notch and Delta [6, 7, 8].

Afek *et al.* compared statistics derived from the observed SOP selection times with several in silico models for stochastic accumulation of Notch and Delta. They finally constructed a consistent model with stochastic rate change that did not require knowledge about the number of active neighbours and used only threshold (binary) communication. Based on this stochastic rate change model, they proposed a new general-purpose distributed algorithm for solving the MIS selection problem in an arbitrary network [2].

The algorithm proposed by Afek *et al.* is fully synchronous and operates over discrete time steps. At each step, each node may choose, with a certain probability (that varies over time), to signal to all its neighbours that it wishes to join the independent set. If a node chooses to issue this signal, and none of its neighbours choose to do so in the same time step, then it successfully joins the independent set, and becomes inactive, along with all its immediate neighbours. However if any of these neighbouring nodes issue the same signal at the same time step, then the cell does not succeed in joining the independent set at that step. This process is repeated until all nodes eventually become inactive. This algorithm is remarkably simple: it requires no knowledge about the number of active neighbours and uses only one-bit messages. The computation at each individual node can be described by a simple automaton, as shown in Figure 2.

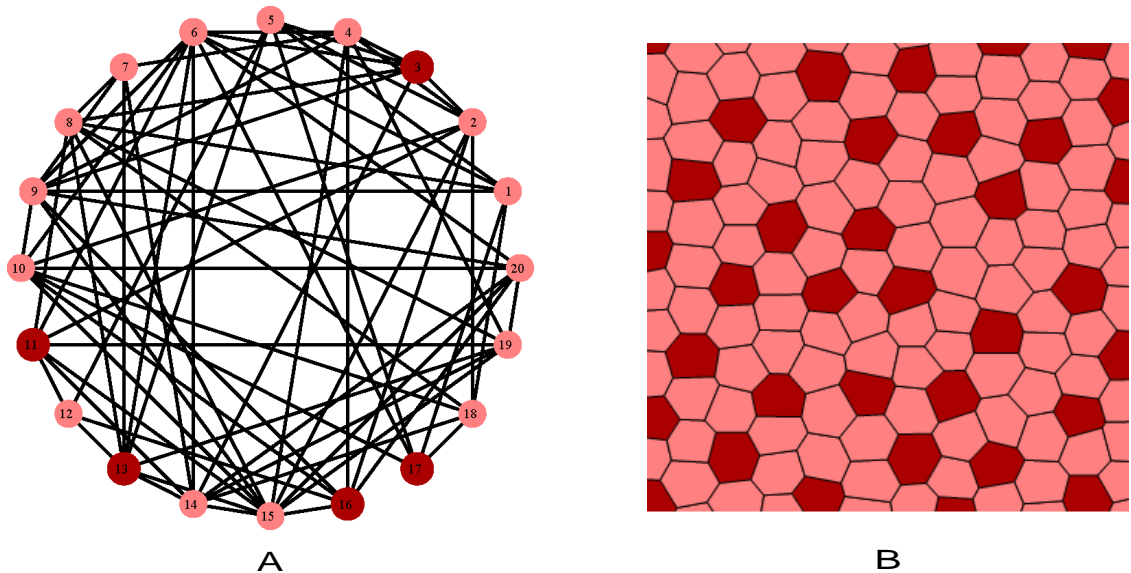

**Figure 1:** (A) An MIS selected from a random undirected graph with 20 nodes. The set of vertices $v_3, v_{11}, v_{13}, v_{16}, v_{17}$ is an MIS because no two nodes in this set are adjacent, and no further node of the graph can be added to this set without violating this property. (B) The selection of SOP cells in the fly appears to be formally similar to an MIS selection proble: each cell either becomes an SOP or a neighbour of an SOP, and no two SOPs are neighbours.

As originally presented [2], the algorithm uses a sequence of gradually increasing global probability values calculated from the total number of nodes of the graph and its maximum degree. The algorithm was further refined by Afek *et al.* in a later paper [1]. In the later version the probability values are chosen according to a fixed pattern, so that the individual nodes require no knowledge about the size of the graph or its maximum degree.

However, in both versions the new approach has a major drawback: the expected number of time steps required is $O(\log^2 n)$. This makes the algorithm usable, but not as fast as previous algorithms. The most well-known distributed algorithm for maximal independent set selection is the elegant randomized algorithm of [3, 16], generally known as Luby's algorithm, which has an expected running time which is $O(\log n)$. Several other previous algorithms have also achieved an upper bound of $O(\log n)$ on the expected number of steps required to compute a maximal independent set, and this has been shown to be the best possible bound that can apply for all networks [18].

To investigate the running time in practice, we implemented the refined version of the algorithm, described in [1], where the probabilities are chosen by repeatedly sweeping across a wider and wider range of different values. Following the scheme specified in [1], we divide the computation into *phases*, numbered $1, 2, 3, \ldots$. Each phase $k$ consists of $k + 1$ time steps. The value of the probability $p$ varies as follows: during each phase $k$ the value of $p$ is 1 initially, and gets halved after each successive time step in phase $k$. Thus, the value of $p$ during the computation will take the following values in successive time steps: $\underline{1, \frac{1}{2}}, \underline{1, \frac{1}{2}, \frac{1}{4}}, \underline{1, \frac{1}{2}, \frac{1}{4}, \frac{1}{8}}, \underline{1, \frac{1}{2}, \frac{1}{4}, \frac{1}{8}, \frac{1}{16}}, \ldots$ (where the underlinings indicate the phases). We ran this version of the algorithm on random networks with different numbers of nodes, where each edge is present with probability $1/2$. We found that the mean number of time steps required to complete the algorithm and choose a maximal independent set in these networks was close to the exact value of $\log^2 n$, where $n$ is the number of nodes (and the logarithm is to base 2). These experimental results are shown as the upper line of data points in Figure 3.

## 2. RESULTS

In this paper we investigate whether the simple algorithmic approach proposed by Afek *et al.* can be improved to make it competitive with the standard approaches in the number of steps required. Our first analytical result gives a negative answer to this question by showing that computing a maximal independent set using this algorithmic approach will require more than some constant multiple of $\log^2 n$ time steps on some families of networks with $n$ nodes *whatever sequence of global probability values is used* (see Theorem 1).

Does this mean that the simple mechanism used by biological cells for "fine-grained" pattern formation has an inher-ently lower efficiency compared with carefully engineered algorithms such as Luby's algorithm? To answer that question we looked more closely at the mechanism of "fine-grained" pattern formation during cell development.

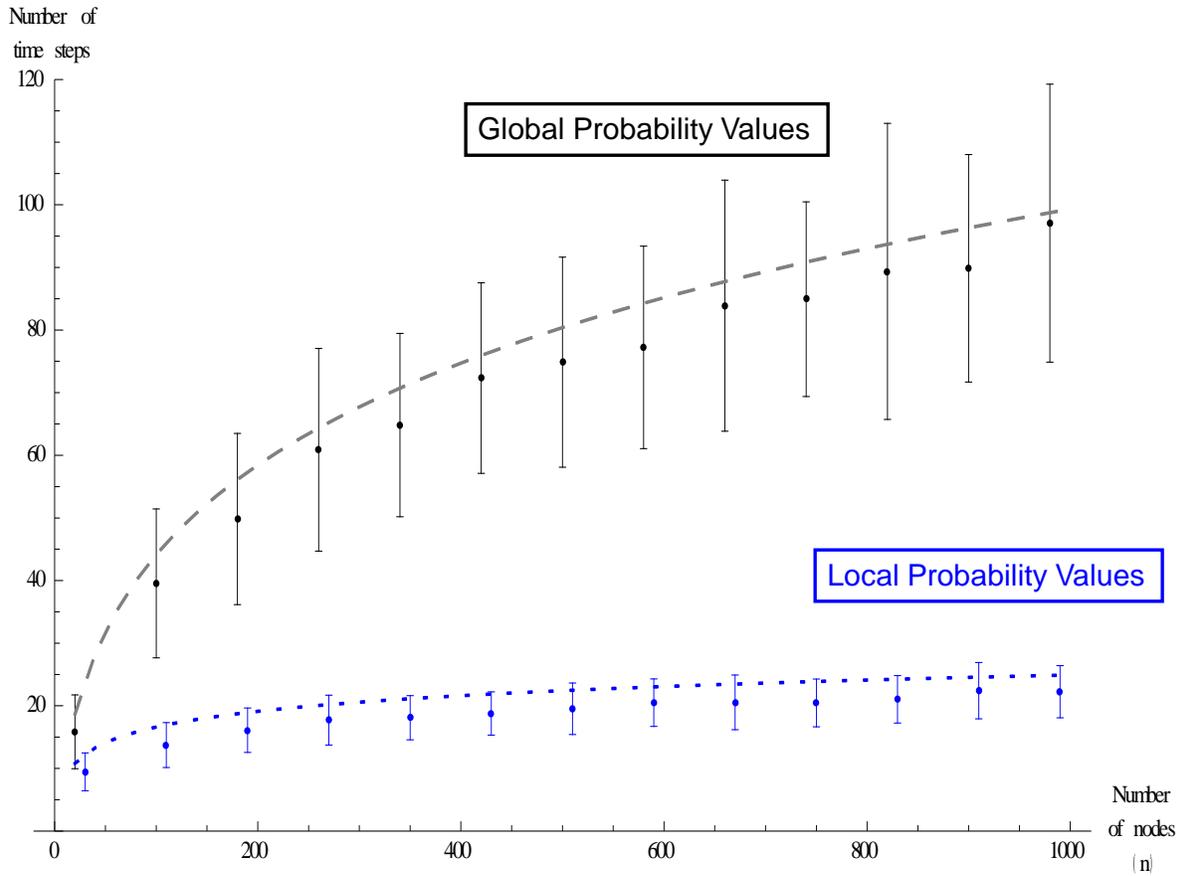

Figure 3: Actual performance of the computation of the MIS on random networks where each edge is present with probability $1/2$. The upper points (black) show the mean number of time steps taken by the algorithm over 100 trials with global sweeping probabilities as specified in [1]. The lower points (blue) the same for locally chosen probabilities with feedback. Error bars indicate standard deviations over 100 trials in each case. The upper dashed line shows the value of $\log^2 n$ and the lower dotted line shows the value of $2.5 \log n$ for comparison (all logarithms to base 2).

| | |
|---|---|
| 1. | Set local value of $p = 1/2$; |
| 2. | **while** active, at each time step **do** |
| 3. |     FIRST EXCHANGE |
| 4. |     With probability $p$, start signalling to all neighbours; |
| 5. |     **if** any neighbour is signalling **then** |
| 6. |         Stop signalling (if started); |
| 7. |         Reduce $p$ |
| 8. |     **else** |
| 9. |         Increase $p$ (up to a maximum of $1/2$) |
| 10. |     SECOND EXCHANGE |
| 11. |     **if** signalling **then** |
| 12. |         Join the MIS; |
| 13. |         Terminate (become inactive) |
| 14. |     **else if** any neighbour is signalling **then** |
| 15. |         Terminate (become inactive) |

Table 1: The algorithm at each node

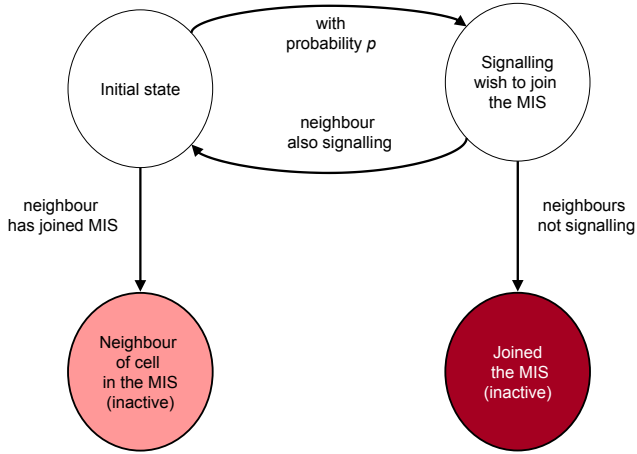

**Figure 2: Abstract state-based description of the process at each node. At each step a node may signal that it wishes to join the independent set by moving to the state at the top right with probability $p$ (which varies with time). It then responds to the signals from neighbouring cells.**

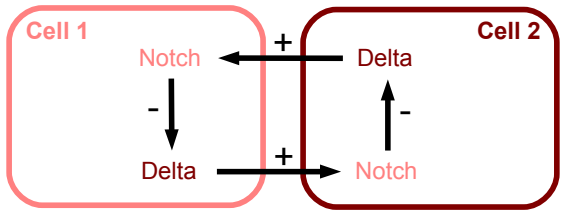

**Figure 4: Notch-Delta signalling constructing a positive feedback to amplify small differences between cells. A slight excess of Delta in cell 2 enables the positive feedback loop that leads to mutually exclusive signalling states of the two cells.**

The Notch-Delta signalling pathway provides a communication channel between neighbouring cells during development. It is thought to play a critical role in the formation of "fine-grained" patterns in the development of many organisms [21, 9], helping to generate distinct cell fates among groups of initially equivalent neighbouring cells. In particular, many studies have shown that Notch-Delta signalling regulates the selection of *Drosophila* neural precursors from groups of equipotent proneural cells in a way which resembles the MIS selection problem [5, 4, 21]. The transmembrane protein Delta has been shown to have two activities: Delta in one cell can bind to, and transactivate, the transmembrane protein Notch in its neighbouring cells [5]; Delta and Notch in the same cell mutually inactivate each other [11, 17]. The interaction of the transmembrane proteins Notch and Delta constitutes a positive feedback mechanism which generates an ultrasensitive switch between two mutually exclusive signalling states: *sending* (high Delta/low Notch) and *receiving* (high Notch/low Delta). A slight excess of Delta production in one cell can generate a strong signalling bias in one direction: the cell becomes a sender and its neighbours become receivers [21]. At the multicellular level, this lateral inhibition mechanism can break the symmetry among cells and amplify small differences between neighbouring cells, thus facilitating pattern formation (see Figure 4).

It is clear that the cells participating in this developmental process do not act autonomously - they continuously adjust their behaviour based on the signals being made by the cells around them. We abstracted from the positive feedback mechanism of Notch-Delta signalling to construct the distributed algorithm described in Table 1. It uses similar processing at each node to the algorithm of Afek et al, but each node now has its own independently updated probability value. These values all start at $1/2$, but are decreased by some fixed factor whenever one or more neighbouring cells signal that they wish to join the independent set. They are also increased by the same factor (up to a maximum of $1/2$) whenever no neighbouring cell issues such a signal.

Our main result shows that choosing the probabilities in this way, using a simple local feedback mechanism, gives an algorithm whose expected time to compute a maximal independent set is only $O(\log n)$ (see Corollary 5). Hence this new simple algorithm performs as well as all previous algorithms for this problem, and, indeed, as well as the theoretical optimal performance.

To illustrate this result we implemented the new algorithm with the probability $p$ at each node varying as follows: $p$ is initially set to $1/2$. At any time step where there is a signal from at least one neighbouring cell the value of $p$ is halved. At all other time steps it is doubled (up to a maximum of $1/2$). We then ran this new algorithm on random networks with different numbers of nodes, where each edge is present with probability $1/2$. We found that the mean number of rounds required to complete the algorithm and choose a maximal independent set in these networks was now close to $2.5 \log n$, where $n$ is the number of nodes (and the logarithm is again to base 2). These experimental results are shown as the lower line of data points in Figure 3.

## 3. LOWER BOUND FOR GLOBALLY CHOSEN PROBABILITY VALUES

Following [1], we refer to the signalling at each node as "beeping". At each time step, each node chooses to beep with a certain probability, and this beep is immediately heard by each of its neighbours.

In this section we consider the class of algorithms described in [2] where each node runs through a preset sequence $p_1, p_2, \ldots$ of probabilities for beeping. We assume that all nodes beep with probability $p_i$ in the $i$th time step. Our first result gives an explicit family of graphs with $O(n)$ vertices, for which *any* such algorithm takes at least a fixed multiple of $\log^2 n$ steps. (We generally omit floors and ceilings for clarity.)

THEOREM 1. *There is a constant $a > 0$ such that the following holds. Let $G$ be the graph consisting of $n^{1/3}$ disjoint copies of the complete graph $K_d$, for each $d = 1, \ldots, n^{1/3}$. Let $T = a \log^2 n$ and let $p_1, \ldots, p_T$ be any sequence of probabilities. Then with high probability, the algorithm using the probability sequence $p_1, \ldots, p_T$ does not terminate within $T$ steps.*

PROOF. Fix $d$, and consider a copy $K$ of $K_d$. The probability that some vertex of $K$ is added to the independent set at the $i$th step is the probability that exactly one vertex of $K$ beeps, and so equals

$$dp_i(1-p_i)^{d-1} \leq dp_i \exp(-(d-1)p_i). \quad (1)$$

Note that the function $xe^{-x}$ is bounded on $[0, \infty)$, and has maximum $1/e$ (at $x = 1$). So for $d > 2$,

$$dp_i \exp(-(d-1)p_i) = \frac{d}{d-1} \cdot (d-1)p_i \exp(-(d-1)p_i) \leq \frac{3}{2e}.$$

Also, for $x \in [0, 3/2e]$, we have $1 - x \geq \exp(-2x)$. So, by inequality 1, the probability that all the vertices of $K$ are still active after $T$ steps is at least

$$\prod_{i=1}^{T} \left(1 - dp_i e^{-(d-1)p_i}\right) \geq \prod_{i=1}^{T} \exp(-2dp_i e^{-(d-1)p_i})$$

$$= \exp\left(-\sum_{i=1}^{T} 2dp_i e^{-(d-1)p_i}\right)$$

$$\geq \exp\left(-\sum_{i=1}^{T} 6dp_i e^{-dp_i}\right).$$

The last inequality follows from the fact that $e^{p_i} \leq e \leq 3$.

Hence if $\sum_{i=1}^{T} 6dp_i e^{-dp_i} < \frac{1}{4} \log n$ then the nodes of $K$ remain active with probability at least $n^{-1/4}$. In that case the probability that the nodes in all the copies of $K_d$ become inactive is at most

$$(1 - n^{-1/4})^{n^{1/3}} \leq \exp(-n^{1/12}),$$

and so with high probability (i.e, tending to 1 as $n \to \infty$) the algorithm fails to terminate.

It follows that we may assume that $\sum_{i=1}^{T} 6dp_i e^{-dp_i} > \frac{1}{4} \log n$ for every choice of $d \geq 3$. We will show that this implies $T = \Omega(\log^2 n)$.

Let us choose $d$ at random. We define a probability distribution for $d$ by

$$\mathbb{P}[d = r] = \frac{c}{r \log n},$$

for $r = 3, \ldots, n^{1/3}$ (where $c$ is a normalizing constant: note that $c = \Theta(1)$, as $\sum_{i=1}^{n^{1/3}} 1/r = \Theta(\log n)$). Then, for any $p \in [0, 1]$,

$$\mathbb{E}[dpe^{-dp}] = \sum_{r=3}^{n^{1/3}} \frac{c}{r \log n} rpe^{-rp} \leq \frac{c}{\log n} \sum_{r=0}^{\infty} pe^{-rp}.$$

But $\sum_{r=0}^{\infty} pe^{-rp} = p/(1 - e^{-p}) < 2$, as $p \in [0, 1]$; so we have $\mathbb{E}[dpe^{-dp}] < 2c/\log n$. By linearity of expectation, choosing a random $d$, we have $\mathbb{E}[\sum_{i=1}^{T} 6dp_i e^{-dp_i}] < 12cT/\log n$.

Hence there is some value of $d$ for which $\sum_{i=1}^{T} 6dp_i e^{-dp_i} < 12cT/\log n$. By the argument above, this quantity must be at least $\frac{1}{4} \log n$, and so we must have $T = \Omega(\log^2 n)$. □

## 4. UPPER BOUND FOR LOCALLY CHOSEN PROBABILITY VALUES

In this section we consider the new class of algorithms described in Table 1, where each node maintains its own independent probability value which varies over time.

*Definition 1.* We define the following distributed algorithm for computing a maximal independent set in an arbitrary graph.

At each time step $t$, there is an integer $n(v, t)$ attached to each vertex $v$, and $v$ beeps with probability $2^{-n(t,v)}$. We set $n(0, v) = 1$ for every $v$.

At each time step, we update according to the following local rules:

- If $v$ beeps and no neighbour of $v$ beeps, then $v$ is added to the independent set and becomes inactive (along with its neighbours).

- If some neighbour $w$ of $v$ beeps, and no neighbour of $w$ beeps, then $v$ becomes inactive (as $w$ is added to the independent set).

- If $v$ does not beep and no neighbour of $v$ beeps, we set $n(t+1, v) = \max\{n(t, v) - 1, 1\}$.

- If some neighbour of $v$ beeps, but no neighbour is added to the independent set, we set $n(t+1, v) = n(t, v) + 1$.

It follows from the analysis of [2] that if this algorithm terminates (i.e., all nodes become inactive) then it correctly identifies an MIS. The only question is the number of time steps required.

Note that, unlike Luby's algorithm [3, 16], it is not true that at every time step we can expect at least some constant fraction of the edges to be incident to nodes that become inactive at that step. For example, in a complete graph nodes will only become inactive when exactly one node beeps. The probability of this happening at the first step is only $n/2^n$, so at the first step the expected number of edges that are incident to nodes that become inactive is only $n^3/2^{n+1}$. Hence we must carry out a more detailed analysis over a sequence of time steps.

THEOREM 2. *There is a constant $K_0$ such that the following holds: For any graph $G$ with $n$ vertices, and any $k \geq 1$, the algorithm defined above terminates in at most $K_0(k+1) \log n$ steps, with probability at least $1 - \epsilon$, where $\epsilon$ is $O(1/n^k)$.*

Before beginning the proof of Theorem 2, it will be useful to define some notation and record a few simple facts.

Let us define a measure $\mu_t(\cdot)$ on $V = V(G)$ by setting

$$\mu_t(v) = \mathbb{P}[v \text{ beeps at time } t] = 2^{-n(t,v)}.$$

(By convention, we set $\mu_t(v) = 0$ if $v$ is inactive at time $t$; this simplifies notation, while allowing us to ignore the contribution of inactive vertices.) For any $S \subseteq V$ we write $\mu_t(S)$ for $\sum_{v \in S} \mu_t(v)$.

We will frequently use the following inequality, which holds for $\delta \in [0, 1]$:

$$1 - \delta \leq \exp(-\delta) \leq 1 - \delta/2. \qquad (2)$$

We will also need the following Chernoff-type inequality: if $X$ is a sum of Bernoulli random variables, and $\mathbb{E}X = \mu$, then for every $\delta > 0$,

$$\mathbb{P}[X > \mu + \delta] \leq \exp(-\delta^2/(2\mu + 2\delta/3)).$$

In particular,

$$\mathbb{P}[X > 2\mu] \leq \exp(-\mu/3). \qquad (3)$$

We will also need the following simple bounds.

PROPOSITION 3. *For any set $S$ of vertices,*

$$\exp(-2\mu_t(S)) \leq \mathbb{P}[\text{no vertex in } S \text{ beeps at time } t]$$
$$\leq \exp(-\mu_t(S)).$$

PROOF. By inequality (2), the probability that no vertex in $S$ beeps at time $t$ is

$$\prod_{v \in S}(1 - \mu_t(v)) \leq \prod_{v \in S} \exp(-\mu_t(v)) = \exp(-\mu_t(S)).$$

On the other hand,

$$\prod_{v \in S}(1 - \mu_t(v)) \geq \prod_{v \in S} \exp(-2\mu_t(v)) = \exp(-2\mu_t(S)),$$

where the last inequality used (2) and the fact that $\mu_t(x) \leq 1/2$ for every $x$ and $t$. □

The set of vertices adjacent to a given vertex $x$ will be called the set of *neighbours* of $x$, and denoted $\Gamma(x)$. We will say that a vertex $x$ at time $t$ is $\lambda$-*light* if $\mu_t(\Gamma(x)) \leq \lambda$, that is, if the total weight of its neighbours is not too large (and so it is not too likely to hear a beep at time $t$). Otherwise we say that $x$ is $\lambda$-*heavy*. Note that a fixed vertex may move back and forth between being heavy and light over time.

Our first result shows that a light vertex is quite likely to be added to the independent set if it beeps.

LEMMA 4. *Let $S$ be a set of $\lambda$-light vertices at time $t$. Then the probability that some vertex in $S$ is added to the independent set at time $t$ is at least $e^{-2\lambda}(1 - e^{-\mu_t(S)})$.*

PROOF. Let us order the vertices of $S$ as $s_1, \ldots, s_r$. Then the probability that some vertex of $S$ is added to the independent set at time $t$ is at least the probability that the earliest vertex of $S$ that beeps is added to the independent set. For $i = 1, \ldots, r$, define events $E_i$ and $F_i$ by

$$E_i = (s_i \text{ beeps}; s_1, \ldots, s_{i-1} \text{ do not beep})$$

$$F_i = (\text{no neighbour of } s_i \text{ beeps}).$$

The events $E_i \cap F_i$ are pairwise disjoint, so the probability that the earliest vertex of $S$ that beeps is added to the independent set is

$$\mathbb{P}[\bigcup_{i=1}^r (E_i \cap F_i)] = \sum_{i=1}^r \mathbb{P}[E_i \cap F_i] = \sum_{i=1}^r \mathbb{P}[E_i]\mathbb{P}[F_i|E_i].$$

It is easily seen that $\mathbb{P}[F_i|E_i] \geq \mathbb{P}[F_i]$, and so by Proposition 3 we have

$$\mathbb{P}[F_i|E_i] \geq \mathbb{P}[F_i] \geq \exp(-2\mu_t(\Gamma(s_i))) \geq \exp(-2\lambda),$$

as $s_i$ is $\lambda$-light, and so

$$\sum_{i=1}^r \mathbb{P}[E_i]\mathbb{P}[F_i|E_i] \geq \exp(-2\lambda) \sum_{i=1}^r \mathbb{P}[E_i].$$

But $\sum_{i=1}^r \mathbb{P}[E_i]$ is simply the probability that some vertex in $S$ beeps, and so by Proposition 3 is at least $1 - \exp(-\mu_t(S))$. Thus the probability that some vertex of $S$ is added to the independent set at time $t$ is at least

$$\exp(-2\lambda) \sum_{i=1}^r \mathbb{P}[E_i] \geq e^{-2\lambda}(1 - e^{-\mu_t(S)}).$$

□

PROOF OF THEOREM 2. Fix a vertex $v$. Let $K_0 = 10^{11}$ and set $K = K_0(k+1)$. Let $\alpha = 10^{-3}$, $\beta = 1/50$ and $\lambda = 7$.

We shall show that, with failure probability $O(1/n^{k+1})$, $v$ becomes inactive within $K \log n = (k+1) \cdot K_0 \log n$ steps. Taking a union bound over all $n$ choices of $v$, it follows that with failure probability $O(1/n^k)$ every vertex becomes inactive and the algorithm terminates within $K \log n$ steps, which proves the theorem.

For $t \geq 1$, let

$$L_t = L_t(v) = \{x \in \Gamma(v) : \mu_t(\Gamma(x)) \leq \lambda\}$$
$$H_t = H_t(v) = \{x \in \Gamma(v) : \mu_t(\Gamma(x)) > \lambda\}.$$

Thus $L_t(v) \cup H_t(v)$ partitions the neighbourhood of $v$ into light and heavy vertices. We will follow the behaviour of $\mu_t(L_t)$ and $\mu_t(H_t)$ over time.

The idea of the argument is roughly as follows: if $\mu_t(L_t)$ is large at many time steps, then by Lemma 4 it is very likely that some neighbour of $v$ will be added to the independent set on one of these occasions, leading to $v$ becoming inactive. If this does not happen, then $\mu_t(L_t)$ will be small most of the time, and we can concentrate on $H_t$. Now vertices that are heavy at time $t$ are likely to hear beeps and so drop in weight (as their beeping probability drops); it will follow that with high probability $\mu_{t+1}(H_t)$ is a constant factor smaller than $\mu_t(H_t)$ most of the time. Now we look at the evolution of $\mu_t(\Gamma(v))$: it may be large and increasing for some small fraction of the time, but mostly it is either shrinking or else it is already small. It will follow that most of the time

$\mu_t(\Gamma(v))$ is small. But then most of the time $v$ will not hear beeps. Since this also implies that $\mu_t(v)$ will be large most of the time, it is very likely that $v$ will beep and not hear any beeps, and so get added to the independent set.

At each time step $t$, we consider the following four possible events:

(E1) $\mu_t(L_t) \geq \alpha$ ['$\Gamma(v)$ has a significant weight of light neighbours']

(E2) $\mu_t(L_t) < \alpha$ and $\mu_t(\Gamma(v)) \leq \beta$ ['$v$ is very light']

(E3) $\mu_t(L_t) < \alpha$, $\mu_t(\Gamma(v)) > \beta$ and $\mu_{t+1}(\Gamma(v)) \leq \frac{1}{\sqrt{2}}\mu_t(\Gamma(v))$ ['the neighbourhood of $v$ shrinks significantly in weight during step $t$']

(E4) $\mu_t(L_t) < \alpha$, $\mu_t(\Gamma(v)) > \beta$ and $\mu_{t+1}(\Gamma(v)) > \frac{1}{\sqrt{2}}\mu_t(\Gamma(v))$ ['the neighbourhood of $v$ does not shrink significantly in weight during step $t$ (and may grow)']

Exactly one of these events must occur at each time step. Note that we know whether (E1) or (E2) occur at the beginning of the time step; if neither occurs, then we must look at the beeps to determine which of (E3) and (E4) occurs.

We organize the proof as a series of claims.

CLAIM 1. *With failure probability $O(1/n^{k+1})$, (E1) occurs at most $(K \log n)/40$ times in the first $K \log n$ time steps.*

Each time that (E1) occurs, it follows from Lemma 4 that some vertex of $L_t$ is added to the independent set (and so $v$ becomes inactive and the process at $v$ terminates) with probability at least $e^{-2\lambda}(1 - e^{-\mu_t(L_t)}) \geq e^{-2\lambda}(1 - e^{-\alpha})$. Let $c_1 = e^{-2\lambda}(1 - e^{-\alpha})$: the probability that (E1) occurs $(K \log n)/40$ times without $v$ becoming inactive is at most $(1 - c_1)^{(K \log n)/40} \leq \exp(-(c_1 K_0/40)(k+1) \log n)$. By our choice of $K_0$, we have $K_0 > 40/c_1$, so this probability is less than $\exp(-(k+1) \log n) = n^{-(k+1)}$. This proves the claim.

The bad event for us will be (E4), so let us bound the probability that (E4) occurs.

CLAIM 2. *At each time step $t$, the probability that (E4) occurs is at most $1/80$.*

If (E4) can occur, then we must have $\mu_t(L_t) < \alpha$ and $\mu_t(\Gamma(v)) > \beta$. If $x \in H_t$ then the probability that no neighbour of $x$ beeps at time $t$ is at most $\exp(-\mu_t(\Gamma(x)) \leq \exp(-\lambda)$. Let $H_t^0$ be the set of vertices in $H_t$ that do not hear a beep at time $t$, and let $H_t^1 = H_t \setminus H_t^0$. Then $\mathbb{E}[\mu_t(H_t^0)] \leq \exp(-\lambda)\mu_t(H_t)$, and so by Markov's inequality

$$\mathbb{P}[\mu_t(H_t^0) \geq 80\exp(-\lambda)\mu_t(H_t)] \leq 1/80. \qquad (4)$$

Now all vertices in $H_t^1$ must halve their weight at the next time step, while vertices in $L_t$ and $H_t^0$ may either halve or double their weight (additionally, some weights may get set to 0 if vertices become inactive). So

$$\mu_{t+1}(\Gamma(v)) \leq \frac{1}{2}\mu_t(H_t^1) + 2\mu_t(H_t^0) + 2\mu_t(L_t)$$
$$= \frac{1}{2}\mu_t(\Gamma(v)) + \frac{3}{2}\mu_t(H_t^0) + \frac{3}{2}\mu_t(L_t)$$

It follows from (4) that, with probability at least $79/80$,

$$\mu_{t+1}(\Gamma(v)) \leq \frac{1}{2}\mu_t(\Gamma(v)) + \frac{3}{2}80e^{-\lambda}\mu_t(H_t) + \frac{3}{2}\mu_t(L_t)$$
$$< \frac{1}{\sqrt{2}}\mu_t(\Gamma(v)),$$

where the final inequality follows from our choices of $\alpha$, $\beta$ and $\lambda$, as $\mu_t(L_t) < \alpha$ and $\mu_t(\Gamma(v)) > \beta$. Thus the probability that (E4) occurs is at most $1/80$. This proves the claim.

CLAIM 3. *With failure probability $O(1/n^{k+1})$, (E4) occurs at most $(K \log n)/40$ times in the first $K \log n$ time steps.*

At each step, the probability of (E4) depends on the past history of the process. However, by Claim 2, it is always at most $1/80$, and so we can couple occurrences of (E4) with a sequence of independent events each occurring with probability $1/80$. It follows that the number of occurrences of (E4) in the first $K \log n$ time steps is stochastically dominated by a binomial random variable $X$ with parameters $K \log n$ and $1/80$. The probability that (E4) occurs more than $(K \log n)/40$ times is therefore, by (3), at most

$$\mathbb{P}[X > 2\mathbb{E}X] \leq \exp(-\mathbb{E}X/3) \leq \exp(-(K \log n)/240)$$

which is $O(n^{-(k+1)})$, proving the claim.

From Claim 1 and Claim 3, we conclude that with failure probability $O(n^{-(k+1)})$, (E1) and (E4) altogether occur at most $K(\log n)/20$ times in the first $K \log n$ time steps. We next show that, with small failure probability, $\mu_t(\Gamma(v))$ is small most of the time.

CLAIM 4. *With failure probability $O(1/n^{k+1})$, $\mu_t(\Gamma(v)) > 2\beta$ for at most $(K \log n)/6$ time steps in the first $K \log n$ time steps.*

Let $T$ be the set of times $t \geq 1$ at which $\mu_t(\Gamma(v)) > 2\beta$. We decompose $T$ into (maximal) intervals of integers, say as $T_1 \cup \cdots \cup T_r$. Let $T_i = [s_i, t_i]$ be one of these intervals. We colour each integer $t \in T_i$ *red* if (E1) or (E4) occurred at the previous step, and *blue* if (E3) occurred (note that (E2) cannot occur, as $\mu_{t-1}(\Gamma(v)) \geq \mu_t(\Gamma(v))/2 > \beta$). Let $r_i$ be the number of red elements and $b_i$ the number of blue elements. We have $\mu_t(\Gamma(v)) \leq \mu_{t-1}(\Gamma(v))/\sqrt{2}$ at blue steps, and $\mu_t(\Gamma(v)) \leq 2\mu_{t-1}(\Gamma(v))$ otherwise. So

$$\mu_{t_i}(\Gamma(v)) \leq \mu_{s_i-1}(\Gamma(v)) \cdot 2^{r_i}/\sqrt{2}^{b_i} \leq \mu_{s_i-1}(\Gamma(v)) \cdot 2^{r_i - \frac{1}{2}b_i}.$$

Since $\mu_{t_i}(\Gamma(v)) > 2\beta$ it follows that

$$r_i > \frac{1}{2}b_i + \log_2 2\beta - \log_2(\mu_{s_i-1}(\Gamma(v))).$$

However, $\mu_{s_i-1}(\Gamma(v)) \leq 2\beta$ in all cases where $s_i > 1$, and $\mu_0(\Gamma(v)) < \frac{1}{2}n$. Summing over $i$, we see that the total number of red elements in $T$ is greater than half the total number of blue elements in $T$ plus $\log_2 2\beta$ minus $\log_2 \frac{1}{2}n$. But red steps correspond to events (E1) and (E4), which altogether occur at most $(K \log n)/20$ times in the first $K \log n$ time steps, so the total number of elements in $T$ is less than $(K \log n)/6$. This proves the claim.

CLAIM 5. *With failure probability $O(1/n^{k+1})$, $v$ hears a beep at most $(K \log n)/3$ times among the first $K \log n$ time steps*

From the previous claim, we may assume that $\mu_t(\Gamma(v)) < 2\beta$ for at least $(5/6)K \log n$ steps out of the first $K \log n$ time steps. If $\mu_t(\Gamma(v)) < 2\beta$, it follows from Proposition 3 that the probability that $v$ hears no beeps is at least

$$\exp(-2\mu_t(\Gamma(v))) \geq \exp(-4\beta) \geq 1 - 4\beta,$$

and so $v$ hears a beep with probability at most $4\beta < 1/12$. Then (3) implies that with failure probability $O(n^{-(k+1)})$ there are at most $(K \log n)/6$ steps among the first $K \log n$ at which $\mu_t(\Gamma(v)) < 2\beta$ and $v$ hears a beep. Since there are at most $(K \log n)/6$ steps at which $\mu_t(\Gamma(v)) \geq 2\beta$, it follows that, with failure probability $O(n^{-(k+1)})$, $v$ hears beeps at most $(K \log n)/3$ times among the first $K \log n$ time steps, proving the claim.

CLAIM 6. *With failure probability $O(1/n^{k+1})$, either $v$ becomes inactive or $\mu_t(v) = \mu_{t+1}(v) = 1/2$ at least $(K \log n)/3$ times in the first $K \log n$ time steps.*

Suppose that $\mu_{t+1}(v) = \mu_t(v)/2$ on $a$ occasions, $\mu_{t+1}(v) = 2\mu_t(v)$ on $b$ occasions and $\mu_{t+1}(v) = \mu_t(v) = 1/2$ on $c$ occasions. We know that $a \leq (K \log n)/3$, by the previous claim, and $b \leq a$ (as $\mu_0(v) = 1/2$, and $\mu_t(v)$ is bounded above by $1/2$), so either $v$ becomes inactive in the first $K \log n$ time steps, or else we must have $c \geq (K \log n)/3$.

We can now complete the proof of Theorem 2. If $\mu_t(v) = 1/2$ and $v$ hears no beeps at time $t$ then $v$ becomes inactive with probability $1/2$. The probability that $v$ remains active for at least $(K \log n)/3$ such steps is at most $(1/2)^{(K \log n)/3} = O(n^{-(k+1)})$. On the other hand, by the claim above, with failure probability $O(1/n^{k+1})$, either $v$ becomes inactive or there are at least $K(\log n)/3$ steps in the first $K \log n$ at which $\mu_t(v) = \mu_{t+1}(v) = 1/2$, and so in particular $\mu_t(v) = 1/2$ and $v$ hears no beeps. We conclude that $v$ becomes inactive with failure probability $O(n^{-(k+1)})$. □

COROLLARY 5. *The expected number of steps taken by the algorithm in Definition 1 on any graph with $n$ nodes is $O(\log n)$.*

PROOF. Let $T$ be the total number of steps taken by the algorithm and let $T' = \lceil T/(K_0 \log n) \rceil$, where $K_0$ is the constant identified in Theorem 2.

By Theorem 2, we have that, for any $k \geq 1$, $\mathbb{P}[T' > k+1] \leq c/n^k$ for some constant $c$. Hence $\mathbb{P}[T' = k+2] \leq c/n^k$, so $\mathbb{E}[T'] \leq 1 + 2 + \sum_{i=3}^{\infty} ic/n^{i-2} = c'$, for some constant $c'$. Hence $\mathbb{E}[T]$ is $O(\log n)$. □

Theorem 2 and Corollary 5 establish upper bounds on the running time that are logarithmic in the number of nodes. Our simulations show that in practice the constants are rather low, leading to an efficient practical algorithm (see, for example, Figure 3).

## 5. BIT COMPLEXITY

We have shown that the expected number of time steps to complete the execution of the randomised algorithm defined in Section 4 grows at most logarithmically with the number of nodes. Another important resource to be considered in any distributed algorithm is the total number of messages sent (the message complexity), and their total size (in bits), which is sometimes referred to as the bit complexity. In this section we will show that the expected number of times that each node beeps is bounded by a constant. Hence the expected bit complexity per channel for this algorithm *does not increase at all* with the number of nodes.

THEOREM 6. *The expected total number of beeps emitted by any node executing the algorithm in Definition 1 is $O(1)$.*

PROOF. Let $v$ be a node executing the algorithm in Definition 1, and consider the whole sequence of time steps until $v$ becomes inactive. At each step, the probability that $v$ beeps is $2^{-n(v,t)}$, which we will denote by $p_t$. The initial value of $p_t$ is $\frac{1}{2}$.

Consider first the subsequence of steps where $v$ hears a beep from its neighbours, halves its own probability of beeping, and reaches a value $p_t$ that is its lowest value so far. The expected number of times that $v$ beeps during this subsequence of steps is $\frac{1}{2} + \frac{1}{4} + \frac{1}{8} + \cdots \leq 1$.

At all other time steps, one of the following 3 things happens:

**Case 1** $v$ hears no beep from its neighbours, and so doubles its probability of beeping (from $p_t$ to $2p_t$)

**Case 2** $v$ hears a beep from its neighbours, and so halves its probability of beeping (from $p_t$ to $\frac{1}{2}p_t$)

**Case 3** $v$ hears no beep from its neighbours, and so leaves its probability unchanged at the maximum value of $\frac{1}{2}$.

Now we consider how many times $v$ beeps during these steps.

If $v$ beeps at any step where Case 3 holds, then it becomes inactive, so the total number of beeps at such steps is at most one.

For each step $t$ where Case 1 holds, and the probability increases from $p_t$ to $2p_t$, we look for a corresponding step $t' > t$ when the probability next drops back to $p_t$. (If there is no such step, because $v$ becomes inactive before the probability returns to $p_t$, then we simply add a dummy step $t'$ to the end of the sequence.) Note that each step where Case 2 holds is now paired with an earlier step where Case 1 holds.

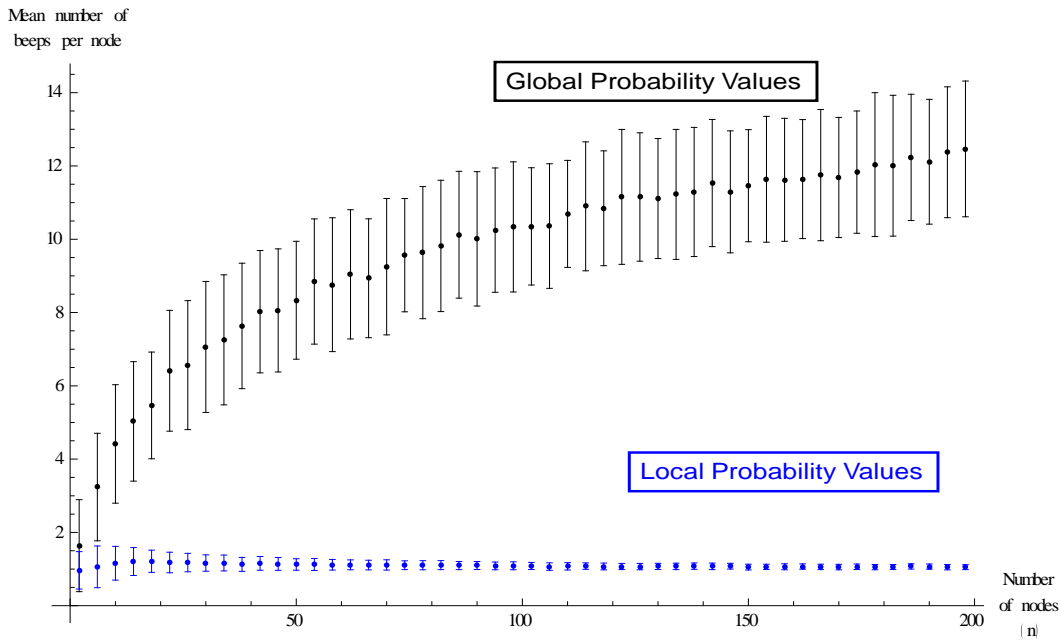

Figure 5: Actual performance of the computation of the MIS on random networks where each edge is present with probability $1/2$. The upper points (black) show the mean number of beeps at each node over 200 trials with global sweeping probabilities as specified in [1]. The lower points (blue) the same for locally chosen probabilities with feedback. Error bars indicate standard deviations over 200 trials in each case.

Now consider the pairs $(t, t')$, and define $B_t$ to be the event that $v$ beeps at either time $t$ or time $t'$ (or both). The total number of times that $v$ beeps at steps where Case 1 or Case 2 holds is at most twice the number of times that events of the form $B_t$ occur.

If $B_t$ occurs, then the probability that $v$ beeps at time $t$ is $p_t/(p_t(1-2p_t) + 2p_t(1-p_t) + 2p_t^2) > \frac{1}{3}$. However, if $v$ beeps at time $t$, then it is added to the independent set, and becomes inactive. Hence the expected number of events $B_t$ that occur before $v$ becomes inactive is less than 3.

We have shown that the expected number of times that $v$ beeps is less than $1 + 1 + 2 \times 3$, which proves the result. □

Our simulations show that in practice the mean number of times that each node beeps is very low and does not increase with the size of the graph. For example, for random graphs with edge probability $\frac{1}{2}$, and for rectangular grid graphs it is around 1.1 (see Figure 5).

Afek *et al.* do not discuss the expected number of beeps at each node in their algorithm with sweeping global probabilities [1]. Our experimental results for this algorithm on random graphs with edge probability $\frac{1}{2}$ are shown in Figure 5, and appear to increase with the size of the network. However, when the initial global probabilities are calculated from the overall network size, and gradually increased, as described in [2], the mean number of beeps at each node does appear to be bounded by a constant. This is consistent with the claim made in [2] that the message complexity for this version of the algorithm is optimal.

## 6. CONCLUSION

In conclusion, we have constructed a simple randomised algorithmic solution to a fundamental distributed computing problem – the maximal independent set selection problem, inspired by the intercellular signalling mechanism used for "fine-grained" pattern formation in many biological organisms. Our algorithm has optimal expected time complexity, and optimal expected message complexity. To achieve this good performance it is necessary to exploit local feedback - we have shown analytically that it cannot be made so time efficient if the processors do not adapt their probability of signalling to their local environment.

Our algorithm uses simple identical processors and one-bit messages, and its expected running time grows only logarithmically with the number of nodes, and is therefore much lower than the time required by a centralised sequential algorithm. Moreover, the individual processors need no information about the global properties of the network, such as its size, and do not need to identify which neighbours sent which message. These features make the algorithm useful for many applications, such as ad hoc sensor networks and wireless communication systems. Selecting a maximal independent set can also be used as a fundamental building block in algorithms for many other problems in distributed computing.

We also note that the algorithm we have proposed here is highly robust, in the sense that it retains its good performance even when various features are changed. For example, the probabilities at each node do not need to increase and decrease by a precise factor - the analysis we have given here can be adapted to a wide range of different values for these factors, which may vary between nodes and over time. Similarly, the initial values for the probabilities at each node may be different from $\frac{1}{2}$, and may vary from node to node, without any significant impact on performance (as long as sufficiently many of them are bounded away from zero). This robustness is likely to be a key feature of the algorithm in any biological context.

The relationship between "fine-grained" pattern formation and the maximal independent set selection problem that we have investigated here is therefore a strikingly successful example of the increasing convergence between systems biology and computational thinking [20].